# Software-Defined Network (SDN) Data Plane Security: Issues, Solutions and Future Directions

Arash Shaghaghi [1,3] · Mohamed Ali Kaafar [2] · Rajkumar Buyya [3] · Sanjay Jha [1]

**Abstract** Software-Defined Network (SDN) radically changes the network architecture by decoupling the network logic from the underlying forwarding devices. This architectural change rejuvenates the network-layer granting centralized management and re-programmability of the networks. From a security perspective, SDN separates security concerns into control and data plane, and this architectural recomposition brings up exciting opportunities and challenges. The overall perception is that SDN capabilities will ultimately result in improved security. However, in its raw form, SDN could potentially make networks more vulnerable to attacks and harder to protect. In this paper, we focus on identifying challenges faced in securing the data plane of SDN - one of the least explored but most critical components of this technology. We formalize this problem space, identify potential attack scenarios while highlighting possible vulnerabilities and establish a set of requirements and challenges to protect the data plane of SDNs. Moreover, we undertake a survey of existing solutions with respect to the identified threats, identifying their limitations and offer future research directions.

Arash Shaghaghi, E-mail: a.shaghaghi@unsw.edu.au
Mohamed Ali Kaafar, E-mail: dali.kaafar@data61.csiro.au
Rajkumar Buyya, E-mail: rbuyya@unimelb.edu.au
Sanjay Jha, E-mail: sanjay.jha@unsw.edu.au

[1] School of Computer Science and Engineering,
The University of New South Wales (UNSW), Australia
[2] Data61, CSIRO, Australia
[3] Cloud Computing and Distributed Systems (CLOUDS) Lab, School of Computing and Information Systems,
The University of Melbourne, Australia

## 1 Introduction

Traditional IP network devices are purpose-built and application-specific with integrated circuits and chips designed to achieve high throughputs. This 'hardware-centric' network model requires network operators to configure each device separately through low-level vendor-specific commands. Hence, in heterogeneous networks, network configuration is tedious, and automatic reconfiguration and response are virtually impossible. Moreover, the data and control plane bundling reduces flexibility, hinders innovation and slows down the evolution of networking infrastructure. In fact, as shown by recent studies, in the long run, traditional networks are incapable of coping with the increasing demand and continuous expansion in the number of devices and applications brought through advances in Cloud Computing, Internet-of-Things (IoT), and Cyber-Physical Systems [43, 62, 71, 77].

Software-Defined Network (SDN) is a new network paradigm that decouples the control from data plane in networking devices. This architectural recomposition places the 'brain' of the network on a specialized central controller, enabling centralized management and global view of the network. The data plane is composed of 'dummy' devices, forwarding packets based on rules specified remotely. These rules may be specified by the application running atop of the controller and triggered according to packet-level extracted information.

SDN's layered architecture follows the 'separation-of-concerns' principle [30], which is a fundamental security engineering requirement and is missing in today's Internet architecture. Hence, in theory, SDN lays a solid ground for improving the security of networks, and tremendous efforts have already been made to leverage the capabilities of SDN to enhance security for both network providers and users. The SDN security literature is split into research aiming to secure software-defined network platform itself, solutions attempting to enhance existing network security services (e.g. firewalls) and proposals on creating new security services. In this paper, we take the first direction and our focus is on the security threats associated with data plane of SDNs.

The roadmap ahead is as follows: we start by presenting an overview of SDN architecture and capabilities, where we highlight the latest advances and trends in SDN data plane research (§2). Thereafter, present a taxonomy of attacks against SDNs based on their scope and impacts (§3). To the best of our knowledge, this categorization does not exist in the literature, which we believe is fundamental for better understanding of different attack impacts in SDNs and develop practical countermeasures. In §4, we formalize the problem of securing SDN's data plane, identify the attack vectors and highlight the underlying vulnerabilities. Afterwards, we establish a set of requirements to protect SDNs against compromised forwarding devices and accordingly, survey existing practical solutions. Finally, we present a set of future research directions in §5 and conclude the paper in §6.

## 2 Software-Defined Network (SDN)

Traditionally, computer networks have been divided into three planes of functionality namely, the management plane, the control plane, and the data plane. In a nutshell, network policies are devised at the management plane and passed to control plane for enforcement and executed at the data plane. Hence, the data plane refers to the network forwarding devices that forward the packets, the control plane represents the protocols used to configure the forwarding tables, and management plane includes the set of software services used to configure the control functionality (e.g., SNMP, NETCONF, etc.). Traditional IP networks, follow a 'hardware-centric' model where the control and data plane are developed and embedded in the same device by the device vendor. The resulting outcome has been quite effective in terms of network performance and resilience. Nevertheless, this architecture is very resistant to change, slow in adopting innovations and quite complicated to setup, troubleshoot and manage.

Software-Defined Network (SDN) has emerged with its largest special envoy being the loose coupling between the control and data plane. Hence, SDN moves away from a vertical integration of network components to a horizontal one and adds distinctive separate functioning layers for policy definition, enforcement and implementation. We present an overview of SDN architecture and its main components in §2.1. Here, we are mostly concerned with SDN's data plane, and in §2.2 we include a more detailed revision of this layer, where we also discuss the recent trends with stateful data planes.

### 2.1 Architecture and Main Components

Software-Defined Network framework facilitates networks programmability and grants the ability to manage, amend and control the network behavior dynamically via open interfaces. It enables centralized control of data plane forwarding devices independent of technology used to connect the devices while maintaining live and centralized network-wide view of all the data path elements. SDN enables long-awaited features such as on-demand resource allocation, self-service provisioning and truly virtualized networking through its intelligent orchestration and provisioning system. The high-level reference SDN architecture promoted by the Open Networking Foundation (ONF) is shown in Figure 1. The architecture is composed of three main layers namely, the Data Plane, Control Plane and Application Plane. Each layer has its own specific functions and the components that are always present in an SDN deployment include the Southbound API, SDN Controller (or Network Operating System), Northbound API and network applications. In the following, we present a succinct overview for each of these components through a bottom-up approach. Understanding the core properties of these components play a role when designing solutions to secure the data plane of SDNs.

#### 2.1.1 Data Plane

The data plane is composed of networking equipments such as switches and routers specialized in packet forwarding. However, unlike traditional networks, these are just simple forwarding elements with no embedded intelligence to take autonomous decisions. These devices communicate through standard OpenFlow interfaces with the controller - which ensures configuration and communication compatibility and interoperability among different devices. An OpenFlow enabled forwarding device has a forwarding table, which is constitute of three parts: 1) Rule matching; 2) Actions to be executed for matching packets; and 3) Counters for matching packet statistics. The rule matching fields include Switch port, Source MAC, Destination MAC, Ethernet Type, VLAN ID, Source IP, Destination IP, TCP Source Port, TCP Destination Port. A flow rule may be defined a combination of these fields. The most common actions include 1) Forward the packet to outgoing port(s); 2) Encapsulate and forward to controller; 3) Drop; 4) Enqueue and 5) Modify Field. The most common case is to install a default rule to instruct the switch to forward the packet to the controller for a decision.

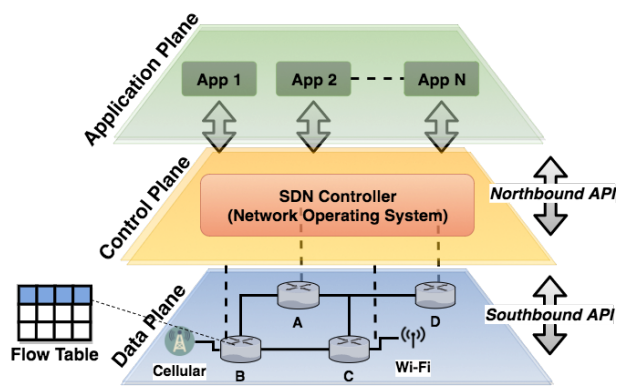

Fig. 1: SDN Architecture

SDN has enabled the introduction of Software switches, which are deemed to be a promising solutions for data centers and virtualized network infrastructures [7, 9]. These have been very attractive to data center networks with the most dominant examples being Open vSwitch [6], Switch Light [4], ,Pica8 [8], Pantou [2] and XorPlus [10].

#### 2.1.2 Southbound API

The southbound API is one of the most critical components of an SDN system, which bridges in-between forwarding devices and the control plane. It enables the controller to control the network behaviour by managing flow entries of all underlying switches. The southbound API provides a common interface for the upper layers enabling the controller to use different southbound APIs (e.g., OpenFlow, POF [102], OpFlex [101] and OpenState [29]) and protocol plug-ins to manage existing or new physical or virtual devices (e.g., SNMP, BGP, and NetConf). These are essential both for backward compatibility and heterogeneity. Therefore, on the data plane, a mix of physical devices, virtual devices (e.g., Open vSwitch, vRouter [100]) and a variety of device interfaces (e.g., OpenFlow, OVSDB [85], OF-Config (OpenFlow Configuration and Management Protocol), NetConf, and SNMP) can coexist.

Currently, OpenFlow is the most accepted standard for southbound standard. The OpenFlow protocol provides three main type of information to the Network Operating System (NOS): 1) Packet-In message: whenever a forwarding device does not have a matching flow rule for a packet or there is an explicit rule for a packet specifying this; 2) Event-based messages: each time a link or port change is triggered; and 3) Flow statistics generated by the forwarding devices.

#### 2.1.3 SDN Controller

The 'brain' of the network, which generates network configurations based on the policies defined by the network operator. It abstracts the lower-level details and makes them available to the application plane through essential services (e.g., network topology, state, device discovery, and etc.) and common APIs to developers.

A diverse set of controllers are available each with their own design and architecture. One of the most prevailing factors in differentiating available controllers is as to whether the controller has a centralized or distributed architecture. With a centralized controller, a single entity is responsible to manage all of the forwarding devices. This architecture has two main limitations: 1) Single point of failure threat; and 2) Scaling limitations. The best known centralized controllers that can achieve the level of throughput required by data center networks include NOX-MT [105], Maestro [79], Beacon [38], Ryu NOS [3] and Floodlight [4]. These NOSs employ multi-threaded designs deploying parallel multicore architectures and achieve processing capabilities such as to deal with up to 12 million flows per second [38]. Distributed controllers such as ONOS [28], Onix [59], HyperFlow [104], PANE [40] and DISCO [86] provide much better scaling support and are much more resilient to logical and physical failures by spreading independent controllers across different network segments. Distributed controllers are equipped with East and Westbound APIs allowing controllers to exchange data, algorithms for consistency models and monitoring information. SDNi [112] is an attempt to standardize the East and Westbound interface.

A typical SDN controller provides a set of core functionalities including topology manager, stat manager, device manager, notification manager, shortest path forwarding and security mechanisms. The first four components are self-descriptive in a networked environment and the security mechanisms provide services such as isolation and security enforcement between services and application (e.g., rules generation by low priority services do not overwrite rules created by high priority applications).

*2.1.4 Northbound API*

Along with the southbound API, the northbound API is one of the key abstractions of SDN. The northbound API is a software ecosystem providing a common interface for developing applications. It abstracts the low-level instruction sets used by southbound interfaces to program the forwarding devices. This application tier often includes global automation and data management applications, as well as providing basic network functions such as data path computation, routing, and security.

Controllers offer quite a broad variety of northbound APIs such as ad hoc and RESTful APIS, multilevel programming interfaces and file systems. However, as of today, there is no standardization for the northbound API yet. In fact, existing controllers such as Floodlight, Onix, and OpenDaylight [73] implement their own northbound API with different specifications and programming languages. Moreover, SDN programming languages such as Frenetic [41], Nettle [108], NetCore [74], Procera [109], and Pyretic [75] also have their own specification and customisation of the northbound API. The lack of a standard API is likely due to the varied nature of applications that can be installed at the Application Plane (see §2.1.5), which can include network virtualization schemes, managing cloud computing systems, security application and other disparate or specialized functions. As an example, work on open northbound APIs is being done for specific vertical applications. OpenStack has developed the Quantum API, which is a vendor-agnostic API for defining logical networks and related network-based services for cloud-based systems. Several vendors have developed plugins for Quantum, which has helped it to become the default networking API for OpenStack, one of the largest open source cloud management platforms [57].

*2.1.5 Application Plane*

The network applications dictate the behavior of the forwarding devices. The applications submit high-level policies to the control plane, which is responsible for enforcing these policies by implementing them as flow rules on network forwarding devices. An SDN application plane consists of one or more network applications (e.g. security, visualization etc.) that interact with controller(s) to utilize abstract view of the network for their internal decision making processes. An SDN application installed atop of the controller is comprised of an SDN App Logic and A-CPI Driver (used to communicate with the controller) [50].

Many different SDN applications have already been developed, and the the current focus is to have an App Store support for SDNs [62], where customers can dynamically download and install network apps. Most SDN applications fall into one of the following five categories: traffic engineering, mobility, and wireless, measurement and monitoring, security and dependability and data center networking. There is a category of applications that leverage the SDN capabilities to build solutions that did not exist before. For instance, Policy Enforcement as a Service (PEPS) [96] provides inter-layer and inter-domain access control enabling a 'defense in depth' protection model, where unsuccessful access requests are dropped before engaging the data provider server. This could potentially be used to build the next generation of firewalls and improve perimeter security against Denial of Service (DoS) attacks. Alternatively, solutions such as AWESoME [106] leverage SDN capabilities to introduce the novel concept of 'per service' management allow administrators to manage all traffic of a service comprehensively, i.e., steering all traffic generated by the user accessing a given service, and not just the traffic related to first-party servers.

SDN Application plane is an ongoing area of research and surveying papers in this area is not within the scope of this paper. Here, we provide an overview of security and dependability applications in §3, and we refer the interested reader to [62] for a detailed survey of the existing well-known network applications.

2.2 Latest Advances in SDN Data Plane

The original OpenFlow standard was too restrictive and various proposals have emerged adding extra flexibility to it. They can be categorized in three different directions including A) adding multiple flow tables to forwarding devices, B) improving the match rule flexibility and C) stateful data planes. While our focus is on the third-case (i.e., stateful SDN data planes), we include a summary of research in the first two cases as well.

*2.2.1 Adding multiple flow tables to forwarding devices*

The single flow table model proposed in original OpenFlow would cause two main problems. First, it would force the developer to deploy OpenFlow rules that combine the Cartesian product of all the required matches. Other than complexity, the exponential increase in the deployed rules would be an extremely inefficient use of flow table capacity, which is typically implemented through the constrained TCAM memory of forwarding devices. Secondly, many applications would benefit from two-stage processing model [42], where packets are first tagged based on some packet characteristics and then matched with rules.

An OpenFlow forwarding devices supporting multiple flow tables (introduced since OpenFlow 1.0 [45]), consist of flow tables composed as a pipeline, where a path through a sequence of flow tables defines how packets should be handled. Proposals have emerged developing advanced hardware solutions enabling a more flexible multiple matching. For instance, Reconfigurable Match Table (RMT) [31], permits flexible mapping, an arbitrary number of match tables having different widths and depths and even supporting matching rules using parameters computers from previous matches.

*2.2.2 Improving the match rule flexibility*

The first OpenFlow version just matched 12 fields. However, greater matching flexibility was recognized as different network applications at different layers may require the capability to associate actions to different types of matches. Hence, the recent OpenFlow versions add support for more than 40 fields including finer-grained fields such as TCP flags.

*2.2.3 Stateful Data Planes*

In order to reduce the switch-to-controller signaling and the associated latency issues, recently there have been proposals to introduce some very specific stateful operations into the data plane forwarding devices. One of the most motivational use-cases for this emerging trend is link failure. With the original OpenFlow specification, when a data link fails, the forwarding device should seek instructions from the controller to setup rectifying rules (e.g., forward all traffic to a backup link). In this case, there is a small interval of time when all packets will be lost, and in large networks, the resulting number can be quite large. For instance, for a 100ms response time from the controller on 10 Gbps link would cause up to 30 million packets to be lost. Recent versions of OpenFlow specification introduce optional support for some stateful capabilities - e.g., fast failover, select group type, and synchronized tables for learning type functionalities are available in OpenFlow version 1.5.1 [1].

Up to this date, three proposals have emerged for stateful SDN data planes, which go beyond the basic features introduced with OpenFlow and that require architectural upgrades to be deployed. In general, stateful SDN data plane proposals have three basic principles in common: 1) retaining state information of the flows within forwarding devices, 2) support programmable packet-level state transition in forwarding devices, 3) granting the forwarding devices the permission to update forwarding states based on flow's local state information without requiring them to seek instructions from the SDN controller.

*OpenState [29]* introduces programmability to SDN's data plane through a special case of eXtended Finite State Machines (XSFM). Each forwarding device keeps two separate tables: State Table and XFSM Table (see Figure 2). The former, stores the current state of the flow based on packets received, which are relevant to that flow. The latter, however, is used to define the rules based on packet's received information. XFSM is modeled as $(S, I, O, T)$, where S is a finite set of states including the start state $S_0$; I is a finite set of events (inputs); O is a finite set of actions (outputs); and T is a rule (state of transition) that maps ⟨state, event⟩ to ⟨state, action⟩. With OpenState, packet processing in a forwarding device is a three-step process. First, the packet's current state is retrieved from the state table and appended to the packet as a metadata field. If, however, the state table lookup retrieves no match, then the forwarding device assigns 'default' state to the incoming packet. Thereafter, the forwarding queries the XFSM table to find the matching rule with ⟨state, event⟩ pair, executed the associated action and updates the state field of the packet as per the 'next-state' field, which is pre-defined in the XFSM table. Finally, the forwarding device's state table is updated based on the 'next-state' value retrieved in the previous step for the corresponding flow.

Similar to OpenState, *FAST [76]* also stores pre-installed state machines inside each forwarding device. In FAST, however, each forwarding device could have several instances of the state machine and each is dedicated to a special application. Moreover, instead of two tables used in OpenState, the data plane implements four tables including 1) State Machine Filter, 2) State Table, 3) State Transition Table, and 4) Action Table (see Figure 2). There is one state machine filter table for all the instances of the state machine, while each state machine has its own state table, state transition

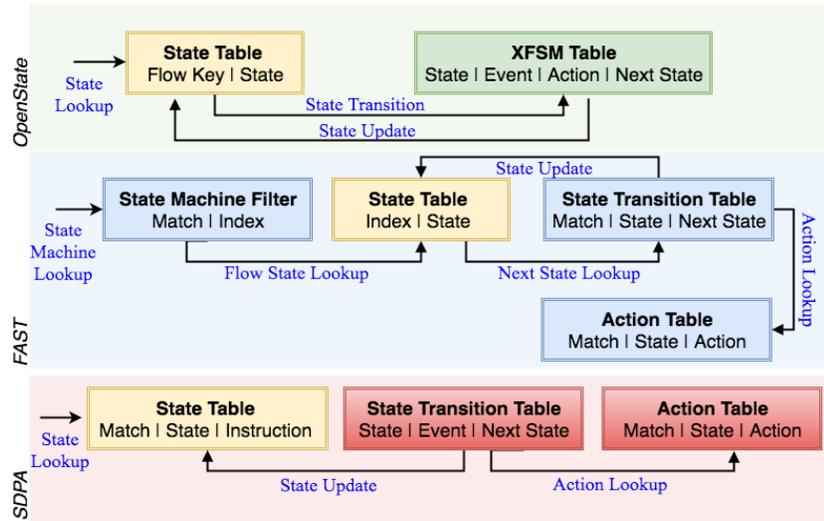

Fig. 2: Tables used in OpenState, FAST and SDPA architectures. Each table is represented as a rectangle containing the table name and the corresponding table columns separated with '|'.

table, and action table. The state machine filter table is used to select the corresponding state machine related to a packet. The state table is a hash table mapping each packet header to a flow state, where each state is stored inside a variable along with its current value - simply put, the state table stores the state information of flows. The forward device uses the state transition table to identify the 'next-state' of a packet based on its current state and packet fields. Lastly, the action table specifies the actions that the forwarding device should execute on the packet based on the packet header and its new state.

FAST also upgrades the control plane of an SDN by introducing two new components: 1) a compiler and 2) forwarding device agents. The former being an offline component responsible to translate the state machines into forwarding device agents. The latter, however, are online components, which 1) manage the sate machines inside the forwarding devices, 2) perform certain local computations based on the updates received from the forwarding device, 3) manage memory restrictions for confined switches through partial implementation of the state machine inside the forwarding device, and 4) handle communication between the forwarding device and the controllers while updating the controller about the local status of the switch.

Another proposal is *SDPA* [120], which is composed of three main tables including State Table, State Transition Table, and Action Table (see Figure 2), as well as a state processing module called Forwarding Processor (FP). The state table has three field values: Match, State, and Instruction. The 'Match' value could be any combination of the header fields of the packet, 'State' value is the flow's current state, and 'Instruction' value may be specified either for a 'state' or 'a packet'. Here, the SDN controller communicates with the FP module and initiates the state tables. Moreover, the controller maintains full control and updated information of the state tables via communications with FP either through periodic or specific event updates. With SDPA, when the first packet of a flow is received, the forwarding device sends the packet to the controller, which determines the state table that the corresponding flow state should be stored in. The rest of the packets pertaining to that flow will be processed locally inside the forwarding device without the controller's intervention.

Several different type of applications have already been built atop of the aforementioned stateful data plane solutions including stateful failure recovery [], HULA [51], port knocking [29], SDN tunneling stateful detection [23] and UDP flooding stateful mitigation[23]. While reviewing each of these solutions is out of scope in this work, we refer the interested reader to [35] for a survey.

## 3 SDN Security Analysis

### 3.1 SDN's Vulnerabilities

Compared to traditional networks, five characteristics of a software-defined network can have the most impact on its security given that they potentially add to the number of vulnerabilities. As illustrated in Figure 3, these characteristics include a centralized controller, open programmable interfaces, forwarding device management protocol, third-party network services and virtualized logical networks. We provide a succinct summary of these characteristics before analyzing the security of SDNs.

*Centralized Controller:* As discussed in §2.1, SDN provides a logically centralized control plane to the network. The controller of the network maintains a global view of the network and programs forwarding devices as per the policies defined at the application layer. While initially controllers were developed as single devices, recently there has been a shift of trend to distributed controllers with the goal of adjusting to scalability and reliability requirements of real-world scenarios. In this case, each set of forwarding devices is assigned to a specific instance of controllers and the controllers, follow a Master/Slave deployment model.

*Open Programmable Interfaces:* In SDNs, there are three main programmable interfaces: A) application plane to control plane, B) east and westbound in the control plane and C) control plane to data plane. Compared to traditional networks, these open interfaces are what make an SDN programmable. A, which also known as Northbound API (see §2.1-IV), enables SDN applications to submit policies to the control plane of the network - e.g., REST APIs. B, is an interface allowing communication among different inter-connected controllers, which may or may not be running in the same domain. C is the southbound API (see §2.1.II), which is the most developed and discussed interface in SDNs up to this date. OpenFlow is the agreed standard for the controller to data plane communications, which allows a controller to program a forwarding device irrespective of the underlying hardware or software in the controller or the forwarding device.

*Forwarding Device Management Protocol:* The forwarding device management protocol along with OpenFlow enables configuration and management of programmable forwarding devices. For instance, the OF-Config protocol may be used to configure an OpenFlow enabled device as well as multiple logical forwarding devices that may be initiated on top of that device. Another example of this protocol includes the OVSDB [85].

*Third-party Network Services:* As an operating system, an SDN controller supports the installation and execution of third-party network services. This allows easy customization, development and innovation, and reduced costs of proprietary services. Third-party services may communicate with a controller either via internal APIs or open Northbound APIs. Moreover, depending on the controller used, applications may be compiled as part of the controller module (e.g., NOX and POX) or may be instantiated at run-time (e.g., OpenDayLight).

*Virtualized Logical Networks:* Network Function Virtualization (NFV), created by a consortium of service providers, is tightly coupled with software-defined networks but it does not depend on SDN for its existence. In a nutshell, NFV virtualizes network services, which were previously hardware-based. It focuses on optimizing network services themselves by decoupling network functions (e.g., DNS, caching, and etc).

While independent, the combination of SDN and NFV leads to a greater value and potential. In fact, in many cases, SDN is linked to server virtualization by enabling multiple logical forwarding devices being instantiated over a shared physical device. This potential has already been explored in the literature, and various proposals have emerged [43].

3.2 Attack Scopes, Vectors and Impacts

Compared to traditional networks, the SDN architecture separates the definition and storage of network policies from their enforcement and implementation. Accordingly, we categories attacks targeting SDN's five main components (see 2.1) as per their impact on network's policy, enforcement, and implementation. Figure 4 illustrates the three main attack types and their relation to SDN's main components. Here, we are focusing on direct associations and potentially, the main motivations of an adversary when targeting any of the SDN's core components. Indeed, attacks against each of the five main components may indirectly fit into all of the three different attack scopes.

3.3 Implementation Attacks

Attacks targeting the Southbound API and Data Plane components of a software-defined network are categorized under 'Implementation Attack'. Figure 5 shows a taxonomy of the main threat vectors associated with this type of attack.

Three different attacks may be used to compromised a data plane including Device Attack, Protocol Attack, and Side Channel Attack. A *Device Attack* refers to all those attacks, where the adversary aims to exploit software or hardware vulnerabilities of an SDN-capable switch to compromise SDN's data plane. In this case, an attacker may target software bugs (e.g., firmware attacks) or hardware features (e.g., TCAM memory) of a forwarding device. For instance, authors in [119] present an inference attack that using the limited flow

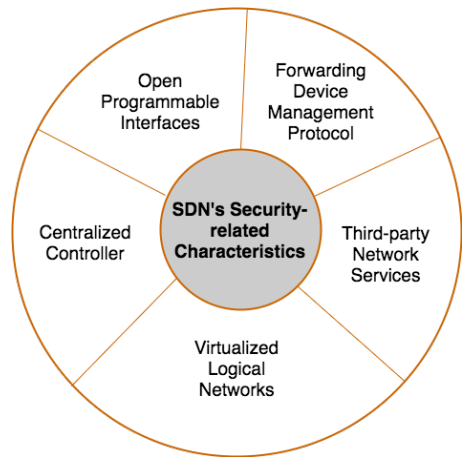

Fig. 3: SDN's Five Main Security-related Characteristics Posing Security Issues

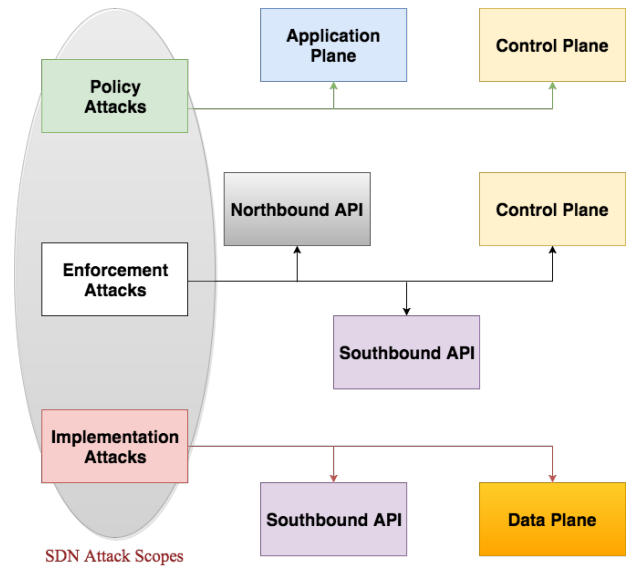

Fig. 4: The Relation Between SDN's Five Main Components and the Attack Scopes

table capacity of OpenFlow switches can infer the network parameter (flow table capacity and flow table usage) with relatively high accuracy. A *Protocol Attack* refers to attacks targeting the data plane of an SDN by exploiting network protocol vulnerabilities in the forwarding devices (e.g., BGP attacks). Authors in [58] provide a detailed study of Denial of Service and information disclosure attacks in SDNs, which are exacerbated due to the nature of OpenFlow. As discussed in [113], most OpenFlow-enabled switch models run custom and independent switch firmware implementations with varying capabilities. For example, the HP 3500yl and 3800 switch models do not support all of the OpenFlow specified 12-tuple match fields in the hardware (TCAM) flow table. This behavior of the switch firmware may be misused to degrade the overall network performance. For instance, the malicious application could install crafted flow rules that override the existing flow rules (IP matching) with hardware-unsupported match fields (MAC matching) specified. A *Side Channel Attack* in this context refers to the case where an attacker may deduce the forwarding policy of the network just by analyzing the performance metrics of a forwarding device. For example, an input buffer may be used to identify rules, and by analyzing the packet processing times, an attacker could identify the forwarding policy [93].

There are four main attacks against the Southbound API of an SDN including Interaction, Eavesdrop, Availability and TCP attacks. While with an *Eavesdrop Attack*, the attacker aims to learn about information exchanged between the control and data plane as part of a larger attack plot, in an *Interception Attack* the attacker's goal is to corrupt the network behavior by modifying the messages being exchanged. For example, authors in [32], present a man-in-the-middle attack using ARP poisoning to intercept the traffic between a client and an SDN controller. Evidently, such attacks could then be expanded to corrupt the network behaviour at a later time. The *Availability Attack* refers to Denial of Service (DoS) attacks, where the Southbound API is flooded with requests causing the network policy implementation to fail. As discussed in [64], attackers can infer flow rules in SDN from probing packets by evaluating the delay time from probing packets and classifying them into classes. Evidently, knowing the reactive rules, attackers can launch DoS attacks by sending numerous rule-matched packets which trigger packet-in packets to overburden the controller.

3.4 Enforcement Attacks

'Enforcement Attacks' aim to prevent a software-defined network to properly instruct when, where and how the policies should be enforced in the network. Hence, attacks targeting the Control Plane, Southbound API and Northbound API may be associated with attacks targeting policy enforcement. Figure 6 illustrates a taxonomy of different attack vectors targeting the enforcement of network policies.

Earlier in §3.3, we denoted different attacks against Southbound API. These attacks may also have an adverse impact when it comes to policy enforcement in an SDN as well. For instance, using a Man-in-the-Middle attack (MITM) attack, an adversary may alter message exchanges such as $Packet\_In^1$ message or $Flow\text{-}mod^2$ and tamper with the controller's understanding of the requirements of the data plane. As well as malicious attacks, the lack of well-defined standards and constant changes in SDN's Southbound API could lead to unwanted, yet malicious involvement in policy enforcement process. For instance, an improperly configured message exchange could lead to invalid or conflicting instructions being set or distributed in the data plane.

---

[1] A *Packet_In* message is sent by a forwarding devices to the controller when a packet does not match any of its flow rules.
[2] A *Flow-mod* message allows the controller to modify the state of an OpenFlow switch.

Authors in [80] analyze the vulnerability of link discovery service in SDN controller showing that the attacker can take advantage of the vulnerability of link discovery service to perform link spoofing attack. The underlying vulnerability behind this attacks is that there is no mechanism built-in SDN controller ensures the integrity/origin of LLDP packets.

Similar to SDN's Southbound API, the Northbound API is susceptible to Interception, Eavesdrop and Availability Attacks. While the nature of both attacks is similar, there are a few key differences: 1) An attacker targeting the Northbound API requires higher-level of access to the system and is potentially sitting on the application plane. There may be cases that the applications do not run on the same device (see [96] for example) and in that case the attack complexity may be reduced as to Southbound API (e.g. where the adversary targets the communication link); 2) The impacts of a compromised Northbound API are potentially larger given that information exchanged between the control and application plane affect network-wide policies. Unlike Southbound API, where OpenFlow is adopted as the standard, the Northbound API lacks any standardization. Specifically, each controller has different specifications for the Northbound API, and this leads to insecure developments. Moreover, a poorly designed Northbound API could also be exploited by malicious applications to manipulate the behavior of other applications through eviction of active sessions, tampering with control message exchanges, and etc.

The third set of attack vectors against policy enforcement originate from the control plane – potentially being the most critical threat against SDNs. Attacks targeting SDN's control plane may be classified into three types: Manipulation Attack, Availability Attack, and Software Hack. *Manipulation Attack* refers to any attempt by an adversary to subvert the controllers understanding of the data plane, which ultimately leads to 'improper' decision making. For instance, an LLDP (Link Layer Discovery Protocol) related as $Packet\_In$ messages may be used to created fabricated links and network topologies. Similarly, an ARP (Address Resolution Protocol) packet relayed as a $Packet\_In$ message could adversely affect the view of the controller. Authors in [46] propose new SDN-specific attack vectors, Host Location Hijacking Attack and Link Fabrication Attack, which can effectively poison the network topology information.

An SDN controller is hosted on a commodity server and may be subject to *Software Hacks* as any other application. For instance, altering a system variable such as time may effectively turn the controller offline. In the case of *Availability Attack*, the adversary aims to make the controller unavailable for a certain period of time for part or all of the network. One way to achieve this is for an attacker to flood the controller with $Packet\_In$ messages - given that these are not authenticated.

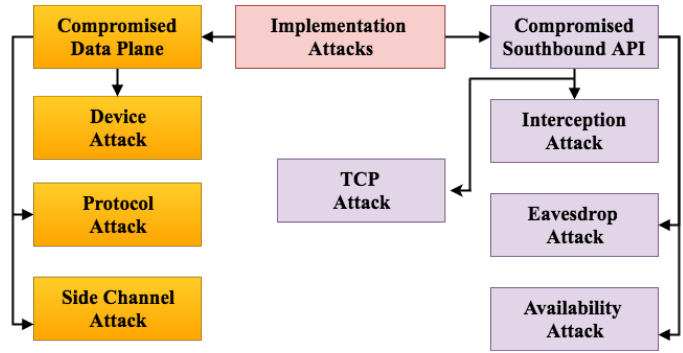

Fig. 5: Taxonomy of Attack Vectors Related to 'Implementation Attacks'

Scalability and configuration conflicts are also vulnerabilities that may be exploited by opportunistic adversaries. An SDN controller is responsible for all decisions in an SDN. Evidently, a single controller will not scale well for large and dense network with a 10 Gbps link network. As discussed in [20], this may be used to deliver attacks such as saturation and single point of failure. Finally, the combination of a single-domain multiple controllers, multi-tenant controllers and multiple OpenFlow architectures may lead to configuration conflicts.

3.5 Policy Attack

'Policy Attacks' refer to the threats targeting SDN's ability to define and store proper network policies. As illustrated in Figure 7, an attacker aiming to target network's policy level aims for compromising SDN's control and application planes. By compromising the controller, an attacker could alter the information shared with applications about the network and compromise their decision making. Potentially, this type of attack may be part of a larger stealthy plot to compromise the network status in the long run or to avoid detection by an intrusion detection system – given that with a compromised controller the attacker has almost full access to the network. In alternative scenarios, the adversary's access to the controller may be restricted to certain functions or period of time motivating an attack to the application plane. Evidently, an 'honest' configuration conflict in the control plane could also be exploited by an adversary to deliver a 'Policy Attack'.

SDN allows the installation of third party apps and as mentioned in §2.1.V, the current goal is to setup an App Store ecosystem for this. Similar to mobile devices, third-party application support adds to the threat vectors against SDNs and ensuring the security and trustworthiness of apps is challenging. As discussed in [92], a major vulnerability that expands the attack surface of compromised applications is that SDN applications are granted complete control and visibility of the network. Hence, a malicious application could use the network state information to manipulate traffic flow for nefarious purposes. [75, 107] further discuss how nested SDN applications pose dangerous threats at this level.

Generally, attacks targeting SDN's application plane may be categorized into: Storage, Control Message, Resource and Access Control attacks.

*Storage Attack:* SDN applications are granted access privilege to shared storage. This access may be exploited to manipulate the internal database targeting the network behavior.

*Control Message Attack:* Control messages exchanged between the control and data plane are fundamental for functioning of an SDN. An arbitrary issued control message by an application might be catastrophic. For instance, a malicious application may take down the network by sending control messages modifying or clearing the flow table entries of switches. For instance, as shown in [97] given that there is no restriction for control messages, an SDN application can issue any control messages at any time. A malicious application continuously generates flow rules to consistently fill up the flow table of the switch and the switch cannot handle more flow rules.

*Resource Attack:* Malicious applications may exhaust expensive and critical system resources including memory and CPU thereby seriously affect the performance of legitimate applications and the controller itself. Moreover, a malicious SDN application may execute system exit command and dismiss controller instances.

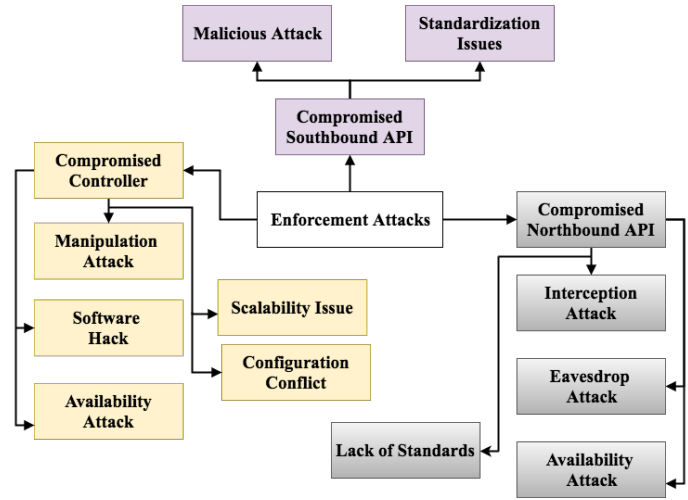

Fig. 6: Taxonomy of Attack Vectors Related to 'Enforcement Attacks'

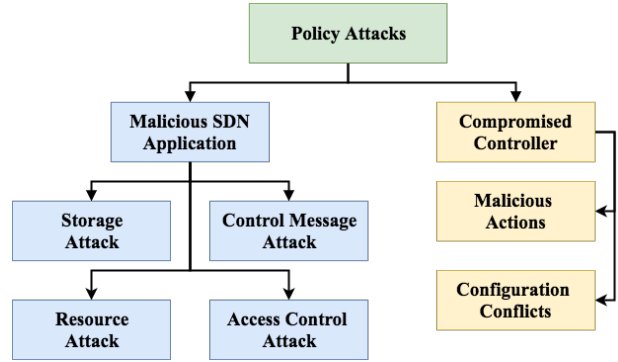

Fig. 7: Taxonomy of Attack Vectors Related to 'Policy Attacks'

*Access Control Attacks:* A common feature among Storage, Control Message and Resource Attacks is access violation. In fact, there is a very limited control in terms of authentication, authorization and accountability in current controllers. We classify all attacks violating required access authorization as Access Control Attack.

3.6 Comparative Analysis

As discussed, attacks targeting an SDN whether falling into Implementation, Enforcement or Policy category can potentially have devastating impacts. Specifically, compared to traditional networks, attack risks have exacerbated given that an attacker could potentially take down a whole network having compromised the main components of an SDN.

Zerkane et al. [116] analyze 114 SDN generic vulnerabilities and compute the severity of theses. They conclude that SDN has a lot of vulnerabilities with high and medium severity because of the weaknesses inherited from classical network architecture and due to its specific characteristics. The vulnerabilities related to access control and those affecting availability are categorized as the most severe. Moreover, they also calculate the impacts of the SDN features on security and identify the control plane components with the highest weight given that SDN architecture is based on the separation, the programmability, and the centralization of the control plane. In contrast, application and network element resource have lower intensities because SDN does not affect their designs.

4 Securing the Data Plane of SDNs

In general, SDN security domain may be split into four main directions. First, research aiming to import existing security services. For instance, [47, 48, 115] aim to design and develop systematic solutions for building reliable firewalls in SDNs. Second, proposals on how to upgrade existing services by leveraging the SDN capabilities. As an example, authors in [114] investigate whether SDN can enhance network security by integrating its capabilities into common security functions. Similarly, solutions such as [18, 82, 111] exploring how SDN can be used to protect networks against malware also fall into this category.

These two directions consisted most of the research in the first few years after SDN's introduction. The recent trend, however, has shifted towards developing innovative security services, which were not feasible to implement before SDN. For instance, using network capabilities to secure Internet-of-Things (IoT) devices (e.g., [81]), smart grid (e.g., [37]), clouds [33] or having inter-layer inter-domain access control [96]. The fourth direction is research aiming to secure the SDN platform itself, which is a critical requirement and has the most direct impact on SDN's adoption. Recently, this has been an active area of research and various solutions have been proposed to secure SDNs at different layers. Proposals such as [49, 88, 89, 98] aim to design and develop secure controllers (see [91] for a categorization). Another category of research aims to secure the northbound interface of an SDN. For example, [92, 110] introduces permissions system that ensures that controller operations are available to trusted applications only. Securing the southbound of an SDN is also a prominent requirement. Authors in [27] provide an overview of the vulnerabilities present in the OpenFlow protocol. Accordingly, solutions such as [21, 54, 60, 99] consider different aspects of OpenFlow that pose security challenges and propose solutions. Authors in [113] provide a comprehensive survey of existing SDN attack studies and evaluate the impact of these attacks against published SDN defense solutions.

In general, the focus on the security of a technology itself is very much driven by the adoption rate as many threats are discovered and addressed as customers deploy it. Fortunately, major industry players such as Google and HP [62] have adopted SDN, and this has boosted research in this area. Several comprehensive surveys have been published summarizing the ongoing efforts in this area including [22, 62, 93, 94]. Here, focusing on SDN's security, we provide an alternative categorization of the SDN security literature (see Figure 8). Out of the four research directions mentioned earlier, all research falling into any of the first three categories is classified as *SDN-based Security Services*. We categories research aiming to secure SDN itself into three groups including research aiming to protect A) SDN's five main components, B) its core features and C) implementations. SDN's core features include centralized management and programmability and various proposals have been developed to ensure these are protected [93]. SDN's implementation includes security different controller platforms, securing the OpenFlow protocol design and implementations, OpenFlow-enabled devices and software forwarding devices such as Open vSwitch. An alternative way of decomposing the literature is according to the SDN components that proposals aim to secure. In §3.2, we discussed different attack vectors targeting SDN's main components as well as their impacts. However, our literature evaluation indicates that one of the least explored areas of SDN security literature is securing its data plane. In fact, even with the latest proposals, there seems to be an oversight regarding the malicious forwarding devices that may exist in SDN data plane [95].

4.1 Why It Matters?

Network forwarding devices have been a very attractive target for attackers. In fact, given the large amount of information that may be exposed through compromised forwarding devices, resourceful adversaries including intelligence agencies have for long aimed to setup backdoors on them. For instance, Edward Snowden uncovered massive investments by NSA to enable large scale surveillance through core network infrastructure [13, 14]. More recently, the 'Vault 7: CIA Hacking Tools' revelations by WikiLeaks [17] disclosed that CIA had actively exploited a common vulnerability in 318 different Cisco routers to carry out surveillance attacks globally. There have also been WikiLeak's revelations on NSA' upgrading labs tampering with forwarding devices before they are released to the market [12]. However, the attack surface against forwarding devices is not limited to resourceful adversaries. Software and hardware vulnerabilities of the devices [15, 24, 34, 65] and vulnerable implementations of network protocols enable attackers to compromise forwarding devices. For instance, as reported in CVE-2014-9295 [16], a novice hacker could execute arbitrary code on routers simply through crafted packets targeting a specific function of the device [11].

A compromised forwarding devices may be used to drop or slow down, clone or deviate, inject or forge network traffic to launch attacks targeting the network operator and its users. For instance, the documents disclosed under 'Vault 7' revelations indicate that compromised routers may have been used for activities such as data collection, exfiltration (e.g. Operation Aurora [5]), manipulation and modification (insertion of HTML code on webpages) and cover tunneling. A compromised routing system may be also used to bypass firewalls and intrusion prevention systems [39], violating isolation requirements in multi-tenant data centers [53], infiltrate VPNs [63] and more.

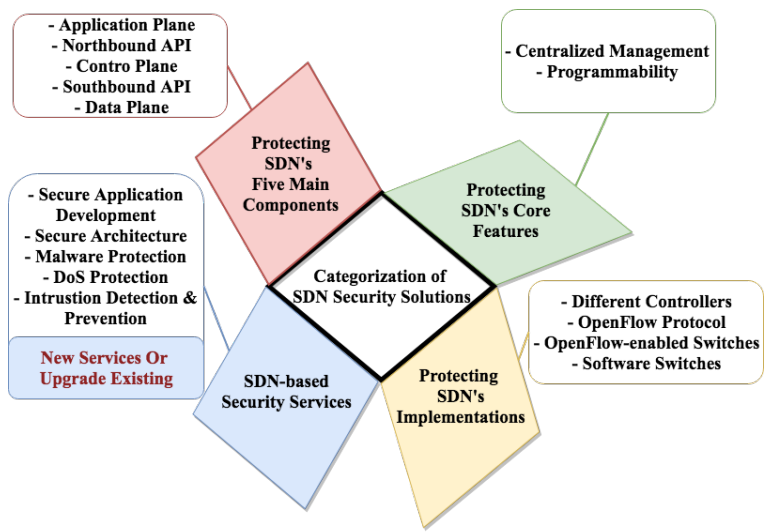

Fig. 8: Categorization of SDN Security Solutions

As discussed in [58, 61], compromised forwarding devices may be used to take down an entire software-defined network. Compared to traditional networks, this shows a compromised forwarding device poses a much higher risk to SDNs. Moreover, as discussed in [36] and [95], SDN also adds to the complications in protecting networks against compromised forwarding devices. This is mainly associated with the following four factors.

- *Incompatibility of the existing solutions:* With the removal of intelligence from the forwarding devices, the defense mechanisms used for traditional networks no longer work in SDNs. In fact, in order to import traditional defenses into SDNs, we would need a fundamental redesign of OpenFlow protocol [72].
- *Unverified reliance of the control plane to the data plane:* SDN controllers rely on *Packet_In* messages for their view of the network. However, as discussed in §3.5, these messages are not authenticated nor verified. A malicious forwarding device may send forged spoofed messages to subvert the controller view of the network – even with TLS authentication in place. The same vulnerability enables a compromised forwarding device the capability to overload the controller with requests launching a Denial of Service (DoS) attack.
- *Software Switches:* Programmable soft-switches such as Open vSwitches run on top of end host servers. Hence, compared to physical switches, soft-switches are more susceptible to attacks with a comparatively larger attack surface.
- *Stateful SDN forwarding devices:* We have already reviewed stateful SDN switches in §2.2.C. In general, adding some intelligence and authority to the data plane has performance advantages such as lower latency response to network events and improved fault tolerance through continuation of basic network operations under failing controllers. Furthermore, well-standardized protocols such as for encryption, MAC learning and codec control message (CCM) exchanges also require some intelligence at data plane. However, these proposals revive some of the vulnerabilities of traditional networks under SDNs.

4.2 Security Implications of Stateful Data Planes

In §3.3, we reviewed attacks targeting the data plane of SDNs and categorized possible attacks into Device Attack, Protocol Attack and Side Channel Attack. Here, we analyze the security implications of the most recent research direction in SDN data planes (i.e. stateful SDN data planes). Our main motivation is to showcase how constant evolution of this technology affect its security and highlight the importance of ongoing research at all levels and for all components of SDNs (see §3).

Compared to standard SDN data planes, three main types of vulnerabilities are added with stateful SDN data planes:

- Unbounded Flow State Memory Allocation: As discussed in §2.2.C, in order to make data planes programmable, each forwarding device must be equipped with memory space to keep track of the state transitions generated by the incoming flows. An attacker may take advantage of the large-in-memory space required for each forwarding device and exhaust the memory of the device.
- Lack of Authentication Mechanisms in the Data Plane: If control independent functionalities were to be implemented in stateful data planes, then this would require the use of probe message between forwarding devices, or information passing between switches and 'piggyback' inside regular traffic packets [35]. Securing inter-forwarding device communications is an important issue, which has almost been ignored in the literature so far. In fact, an attacker may inject fake event/packet into the network impersonating an honest device. Moreover, if the connections are not secured, an attacker may alter the information exchanged between the forwarding devices and change the specific flow states. An attacker could setup fake scenarios, where a link failure has occurred and degrade the network performance.
- Lack of a Central State Management: State inconsistency is an issue of concern for all distributed systems including inter-linked stateful data planes. However, this is more worrisome in this case given that there exists no central entity to manage the synchronization of states inside the forwarding devices. Specifically, as state transition is triggered after packets are received, an attacker has the capability to force state transitions pushing the network into an inconsistent state.

On the other hand, the architectural changes introduced with stateful data planes reduce some of the attacks that were possible in traditional SDNs:

- Enforcement Attacks: We defined Enforcement Attacks in §3.4. Stateful data planes reduce the required communication between control and data plane. This improves the resilience of SDN against Availability Attacks by improving its scalability. As discussed, an attacker could saturate the controller's resources by flooding a large number of spoofed *Packet_In* message targeting both the communication link and controller's resources.
- Implementation Attacks: Stateful SDNs mitigate the following vulnerabilities by design: A) flow information leakage and B) exhaustible TCAM used for flow tables. In A, the attacker learns about the network configuration by passively listening to messages exchanged between the control and data plane focusing on timing information. With stateful data planes, the forwarding devices can be programmed to handle incoming flow without the need to contract the controller. Hence, flow information leakage vulnerability is to a large extend less relevant with stateful data plane deployments. TCAM memory exhaustion is also mitigated by stateful SDN data plane [35].

4.3 Solution Requirements

As discussed in §4.1, securing the data plane of SDNs is even more challenging than traditional network. Adding to this, with ongoing improvements and changes in different SDN components designing a solution to protect networks against compromised data plane forwarding devices is a challenging task. In fact, we believe along with attempts to secure each different piece of data planes, we need a solution capable of automatically detecting compromised forwarding devices and protect networks against them. Hence, any such detection should be based on the forwarding device's main functionality (i.e., packet forwarding) rather than any protocol, software or attack. We posit the 'must-have' features of a working solution to include:

- *Scanning Methodology:* For efficiency and to reduce the detection time, the protection mechanism must systematically and autonomously prioritize forwarding devices for inspection.
- *Distinguish Malicious Actions:* The protection mechanism must be able to distinguish between the specific malicious actions (e.g. packet drop, fabrication, delay, etc.) so that it can be intelligently responded to.
- *Locate Malicious Forwarding Devices:* To effectively protect the network, the protection mechanism must be able to localize the malicious device in the network.
- *Intelligent Response to Threats:* The protection mechanism must be programmable and allow an administrator to customize the response as per the high level network policy and requirements such as Quality of Service. In fact, the data plane is a critical component of the network infrastructure and the proposed solution must not disrupt the network performance during its inspection and analysis stage.
- *Support Stateful Data Plane:* In general, the protection mechanism must be designed with the latest advances in data plane solutions. As discussed in §4.2, stateful data planes are an increasing area of focus and bring along a set of opportunities and challenges.

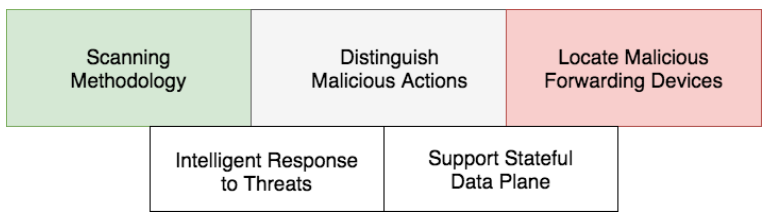

Fig. 9: Required Features For a Working Solution

One of the critical requirements when developing a protection mechanism is to have a clear description of the adversary model. We posit the following adversarial model when developing a working solution:

A resourceful adversary who may have taken full control over one, or all, of the forwarding devices. This is, in fact, the strongest possible adversary that may exist at the SDN data plane. The attacker may drop, replay, misroute, delay, reorder and even generate packets (includes both packet modification and fabrication), in random or selective manner over all or part of the traffic. These capabilities grant the adversary the capability to launch attacks against the network hosts, other forwarding devices or the control plane. For example, executing a Denial of Service (DoS) attack against the control plane by replaying or spoofing *Packet_In* messages.

## 5 Existing Solutions

Existing proposals in SDN data plane security have been known to suffer from inaccurate adversarial model [95]. This limitation directly impacts their adoption and impact. For instance, solutions proposed in [36, 52, 53, 55, 69] assume all, or the majority, of the forwarding devices, are trustworthy. To the best of our knowledge, among existing solutions, only SPHINX and WedgeTail have practical implementations and are evaluated under realistic conditions. Among these two, the only solution that can satisfy the requirements specified in §4.3 is WedgeTail [95]. In the following, we start by reviewing existing solutions aiming to detect packet forwarding anomalies outlining their limitations. Thereafter, we provide an overview of the two most prominent solutions in §5.1 and §5.2 and discuss their limitations in §5.3.

Relevant literature in packet forwarding anomaly detection can be broken down into 1) cryptographic mechanisms, 2) flow statistics, 3) packet probing, and 4) acknowledgement-based mechanisms. Cryptographic mechanisms such as [56, 67, 78, 90, 118] embed signatures in packets and the forwarding devices verify whether the packets have been correctly routed. These approaches suffer from two main limitations for deployment: 1) cryptographic operations incur significant computational overhead and 2) require modification in IP packet formatting. An alternative effective approach to cryptographic solutions is to analyze flow statistics at forwarding device ports (e.g., [25, 68]). However, flow statistic techniques heavily rely on strict time synchronization among forwarding devices, which is hard to achieve in real large-scaled networks and are unable to detect packet modification attacks. Packet probing approaches such as [19, 26, 83] sample and analyze probing packets to detect forwarding anomalies. Majority of these solutions are focused on anomaly detection at first and last hops of a network and result in significant communication overhead. Acknowledgement-based solutions such as [66, 70, 117] detect packet dropping through periodical interaction among neighbouring forwarding devices. In this case, there is also a significant overhead in computation and storage for forwarding device given that each forwarding device should store the entire forwarding path of flows and collects the acknowledgment packets periodically.

### 5.1 SPHINX

Proposed in 2015, SPHINX is a framework to detect attacks on network topology and data plane forwarding. SPHINX is one the very few solutions to secure SDN's data plane that does not assume forwarding devices are trusted. It detects and mitigates attacks originated from malicious forwarding devices through 1) abstracting the network operations with incremental flow graphs and 2) pre-defined security policies specified by its administrator. It also checks for flow consistency throughout a flow path using a similarity index metric, where this metric must be similar for 'good' switches on the path. SPHINX architecture is shown in Figure 10 – the image is imported from author's published paper.

SPHINX leverages the novel abstraction of flow graphs, which closely approximate the actual network operations, to (a) enable incremental validation of all network updates and constraints, thereby verifying network properties in realtime, and (b) detect both known and potentially unknown security threats to network topology and data plane forwarding without compromising on performance. It analyzes specific OpenFlow control messages to learn new network behavior and metadata for both topological and forwarding state, and builds flow graphs for each traffic flow observed in the network. It continuously updates and monitors these flow graphs for permissible changes, and raises alerts if it identifies deviant behavior. SPHINX leverages custom algorithms that incrementally process network updates to determine in realtime if the updates causing deviant behavior should be allowed or not. It also provides a light-weight policy engine that enables administrators to specify expressive policies over network resources and detect security violations.

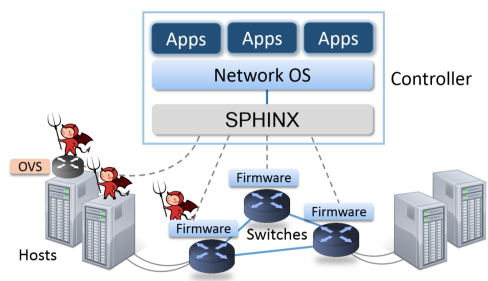

Fig. 10: SPHINX Architecture

5.2 WedgeTail

WedgeTail is a controller-agnostic Intrusion Prevention System (IPS) designed to 'hunt' for forwarding devices failing to process packets as expected. It begins by mapping forwarding devices as points within a geometric space and regarding packets as 'random walkers' among them. WedgeTail tracks packet paths when traversing the network and generates their corresponding trajectories. Thereafter, in order to detect malicious forwarding devices, locate them and identify the specific malicious actions (e.g., packet drop, fabrication, etc.), it compares the actual packet trajectories with the expected ones (i.e., those allowed according to the network policies). If and when a malicious forwarding device is detected, WedgeTail responds as per the administrator-defined policies. For example, an instant isolation policy may be composed of two actions. First, the potentially malicious device is instructed to reset all the flow rules and then, it is re-inspected at various intervals by re-iterating the same packet(s) originally raising suspicion. If the malicious behavior is persistent, the forwarding device may be isolated from the network.

In order to increase the probability of finding malicious devices, WedgeTail begins by prioritizing forwarding devices for inspection. It adopts Unsupervised Trajectory Sampling [84] to cluster forwarding devices into groups of varying priority based on the cumulative frequency of occurrence in packet trajectories - i.e. all of the trajectory database was analyzed. Wedgetail intercepts OpenFlow messages exchanged between the control and data plane and maintains a virtual replica of the network. It uses this virtual replica to compute the expected packet trajectories removing any trust on forwarding devices for this. The expected packet trajectories are computed using Header Space Analysis (HSA) [53] framework. In order to compute the actual packet trajectories, WedgeTail relies on NetSight [44] and queries for the packet history as it traverses the network. However, when NetSight is not available (e.g. in small-sized networks), the packets may be tracked using a custom packet tracking mechanism proposed in the paper.

As shown in Figure 11, WedgeTail is composed of two main parts namely, Detection Engine and Response Engine. The Detection Engine is responsible to retrieve the actual and expected packet trajectories, create the scanning regions and implement the attack detection algorithms. Accordingly, whenever a compromised device is detected, the Response Engine submits policies to the controller to protect the network.

WedgeTail has been tested under simulated environments and with a network composed of more than 300 forwarding devices and 630,000 trajectories it has been capable of locating malicious forwarding devices and identify their specific malicious action in about 90 minutes. For a relatively large network detection of such malicious entity hidden in lowest layer of network infrastructure without defining any policies or manual intervention is quite satisfactory.

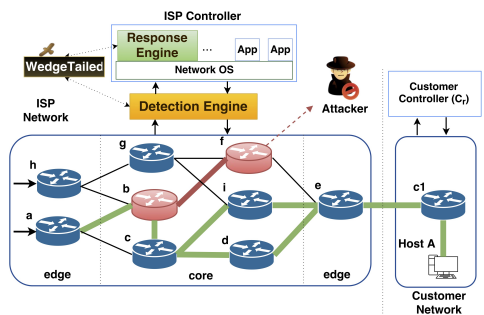

Fig. 11: An abstract representation of WedgeTail over an ISP Network. The red devices are malicious and have been compromised by Attacker. The paths shown in green represent the expected paths for a packet send through forwarding device $fd(a)$ on port $p_i$ to forwarding device $fd(c1)$ on port $p_j$ at Customer Network. The trajectory shown in red depicts a path the same packet took, which is not allowed as per the controller policies.

5.3 Discussion

We argue the following three factors as the main limitation of SPHINX. First, the system does not tolerate Byzantine forwarding faults. In other words, it does not assume malicious forwarding device may behave arbitrarily and therefore, SPHINX is not designed to detect the specific malicious actions performed such as packet drop, fabrication and delay. Second, the detection mechanism mainly relies on the policies defined by its administrator. In fact, the flow-graph component does not validate forwarding device actions against the controller policies but compared to their behavior through time. Hence, if the forwarding device(s) has been malicious since day one it will never be detected or when there are radical network configuration changes SPHINX will have large false positives. Third, SPHINX does not include a scanning regime and has no prioritization when inspecting the data plane for threats. Arguably, an important factor required for optimizing the detection performance in this context. Last but not least, SPHINX requires the majority of forwarding devices to be trustworthy. Although this assumption is realistic, a better solution solution will have to be independent of it.

Essentially, WedgeTail is built on top of the limitations its developers had identified in SPHINX. WedgeTail relies on network snapshots to compute the expected packet behaviors. However, maintaining valid snapshots of large real-world networks may lead to major scalability and performance limitations. WedgeTail's accuracy in detecting attacks over different use-cases and scenarios such as virtualization and VM migrations has not yet been evaluated. Furthermore, WedgeTail's compatibility with distributed controllers has not been addressed and given the shift of the community to such controllers this may be a major barrier in its adoption.

Moreover, neither SPHINX nor WedgeTail are capable of identifying the vulnerability (software, hardware, etc) exploited to compromise a forwarding device.

Both SPHINX and WedgeTail have adopted a proxy model in their prototype implementations. In fact, authors claim that their solution can be imported with improved SDN standardization especially with regard to the controllers. However, further investigation is required to understand SDN forwarding device hardware overhead for such proposals and whether hardware latency will affect the practicality of these solutions in both attack detection and prevention. There is an area of research exploring SDN hardware requirements and adaptations. It is to be explored whether proposals such as [31] or 'server-switch' [87] could make the real-world deployment of solutions such as SPHINX and WedgeTail more relevant.

Recently, Kashyap et al. [103] have proposed teleportation as a new attack against SDNs. Teleportation attack enables an adversary to bypass critical network functions in the data plane, violate security policies and both logical and physical separations [103]. The authors have not evaluated the feasibility of their attack when WedgeTail is in place and provide little abstract discussion how SPHINX may impact the practicality of their attack. Hence, neither SPHINX or WedgeTail have been evaluated against such adversaries and further investigation is required to verify whether such solutions will be capable to protect against such attacks.

6 Future Research Directions

As discussed in §4, SDN data plane security research has been relatively dormant with few proposals emerging only recently. Given the critical role of data plane in the implementation of network policies (see §2.1), extreme value for attackers (see §4.1), increased opportunities for attackers to target them (see §3.2) and ongoing evolution (see §4.2), we regard data plane security as a major challenge for a secure SDN network. Specifically, the fact that compromised forwarding devices can potentially take down an entire SDN system further highlights the importance of the current oversight.

We predict, however, this status will radically change with the increased adoption of SDNs. Specifically, given the potential attacks that a compromised forwarding device may deliver against different technologies that are imported to SDNs such as Software-Defined Clouds [33], we perceive the importance of SDN data plane security to be much more of a focus.

As discussed in §5, the existing solutions suffer from limitations that hinder their real-world application. We expect future research to propose more advanced and practical protection mechanisms on the ground of existing limitations. Specifically, none of the existing solutions have any consideration for the most recent proposals including stateful data planes. Hence, one of the immediate future required extensions is to evaluate the applicability of existing solution for stateful forwarding devices and propose solutions to ensure if any such device is fully, or partially, compromised it can be detected. Another direction of research that we expect to be very active in the future is designing secure hardware for SDN-enabled forwarding devices based the latest software advances and requirements.

Finally, similar to [95], we believe SDN capabilities provide the required ground to develop solutions that secure this technology independent of the underlying software, hardware and protocol. In fact, we believe this is the most reliable approach towards ensuring secure deployment of this open, dynamic and evolving technology.

## 7 Summary and Conclusion

In this paper, we presented a taxonomy of attacks against Software-Defined Networks based on their scope and impacts. Specifically, we discussed how an adversary can leverage vulnerabilities of different SDN components to target network policies, their enforcement, and implementation. Moreover, we argued the importance of security SDN data planes, surveyed the latest advances in data plane development and analyzed the security implications of stateful data planes. Thereafter, we established a set of requirements for a working solution and reviewed the existing solutions with respect to these requirements. Finally, we provided a set of suggestions for future research directions, where we hope to attract researchers' attention to a relatively dormant yet critical aspect of securing SDNs.

**Acknowledgements** We acknowledge the useful comments offered by Sandra Scott-Hayward (Queen's University Belfast) for improving this paper. We also acknowledge the insightful comments provided by the journal reviewers, which helped us to improve the quality of the paper. Arash Shaghaghi acknowledges the Cloud Computing and Distributed Systems Laboratory for hosting his visit at the University of Melbourne, Australia.

Software-Defined Network (SDN) Data Plane Security: Issues, Solutions and Future Directions    21

Software-Defined Network (SDN) Data Plane Security: Issues, Solutions and Future Directions                                    23

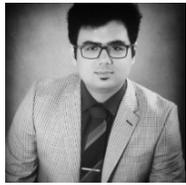
**Arash Shaghaghi** is a PhD Candidate at UNSW Sydney where he is also affiliated with Data61, CSIRO. Previously, he completed an MSc in Information Security at University College London (UCL) and a BSc in Information Technology at Heriot-Watt University of Scotland. Arash has been consistently ranked as Top student in his studies and received multiple awards and scholarships in the past few years. Arash research interests include Network and System Security, Access Control Systems and Applied Cryptography. Arash's cyberhome: `http://Arash.Sydney`

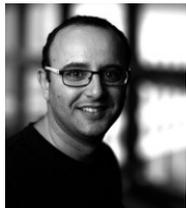
**Mohamed Ali Kaafar** is the group leader of the networks research group at CSIRO Data61, Australia. He was previously a research leader NICTA, Australia, and a research scientist at INRIA RhoneAlpes, France. His research interests include cyber Security, privacy preserving technologies, and networks measurement.

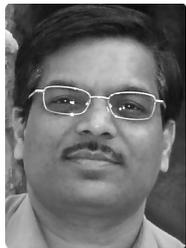
**Rajkumar Buyya** a is Redmond Barry Distinguished Professor of Computer Science and Software Engineering; and Director of the Cloud Computing and Distributed Systems Laboratory at the University of Melbourne, Australia. He is also serving as the founding CEO of Manjrasoft Pty Ltd., a spin-off company of the University, commercializing its innovations in Grid and Cloud Computing. He has authored over 525 publications and seven text books including 'Mastering Cloud Computing' published by McGraw Hill, China Machine Press, and Morgan Kaufmann for Indian, Chinese and international markets respectively. He is one of the highest cited authors in computer science and software engineering worldwide. Recently, Dr. Buyya is recognized as '2016 Web of Science Highly Cited Researcher' by Thomson Reuters. For further information on Dr. Buyya, please visit his cyberhome: http://www.buyya.com.

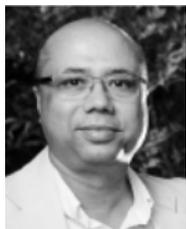
**Sanjay Jha** received the Ph.D. degree from the University of Technology, Sydney, Australia. He is currently a Professor and the Director of the Cyber Security and Privacy Laboratory and the Head of the Networked Systems and Security Group, School of Computer Science and Engineering, the UNSW Australia. He has authored or co-authored over 100 articles in high-quality journals and conferences. His research activities include a wide range of topics in wireless networking and security. He has served on program committees of several conferences. He is a Senior Member of the IEEE Computer Society.